\documentstyle[aps,prl,epsfig,amssymb,balanced,times]{revtex}
\begin{document}
\draft
\title{%\vspace*{5cm}
Hydrodynamic Waves in Regions with Smooth Loss of Convexity of
Isentropes.\\ General Phenomenological Theory}
\author{Michael I. Tribelsky\cite{Trib}$^1$, and Sergei I. Anisimov\cite{Anis}$^2$}
\address{$^1$Department of Applied Physics, Faculty of Engineering,
Fukui University,\\ Bunkyo 3-9-1, Fukui 910-8507, Japan\\ and
Max-Planck-Institut f\"ur Physik komplexer Systeme,\\ N\"othnitzer
Stra\ss e 38, D--01187 Dresden, Germany\\ $^2$Landau Institute for
Theoretical Physics RAN, Kosygin St., Moscow 117334, Russia}
\date{\today}
%Preprint mpi-pks/0009015; Submitted to Phys. Rev. Lett.}
\maketitle
\begin{abstract}
General phenomenological theory of hydrodynamic waves in regions
with smooth loss of convexity of isentropes is developed based on
the fact that for most media these regions in $p\,$-$V$ plane are
anomalously small. Accordingly the waves are usually weak and can
be described in the manner analogous to that for weak shock waves
of compression. The corresponding generalized Burgers equation is
derived and analyzed. The exact solution of the equation for steady
shock waves of rarefaction is obtained and discusses.
\end{abstract}
%
% 47.40.-x Compressible flows; shock and detonation
% 43.25.+y Nonlinear acoustics
% 52.35.Mw Nonlinear waves and nonlinear wave
%
\begin{twocolumns}
Extremely great importance of the so-called elementary wave
patterns, such as shock and rarefaction waves, in hydrodynamics has
been known since the pioneering studies of Riemann\cite{R} carried
out in the beginning of 1860s. The importance is generally
connected with (i) possibility to picture general fluids flows as
nonlinear superposition of elementary waves, propagating as
separate entities and (ii) reflection with the elementary waves the
asymptotic behavior of general solutions of the set of hydrodynamic
equations. For this reason a great deal of attention has been payed
to study the elementary waves, see, e.g., the review
article\cite{MP} and references therein. Nonetheless, a certain
type of the elementary waves, namely shock waves of rarefaction and
sonic waves\cite{MP}, which admits general analytical
consideration, has not been considered in such a manner. The gap is
filled up in the present Letter.

It is well known that the sign of variation of pressure and density
in a weak shock wave is associated with the sign of $(\partial^2
V/\partial p^2)_s$. Usually the sign of the derivative is positive,
and accordingly the shock wave is compressive. However, in some
specific areas of $p\,$-$V$ plane, e.g., close to a critical point
but still in a single-phase region, $(\partial^2 V/\partial p^2)_s$
may become negative. The phenomenon is known as smooth loss of
convexity of isentropes\cite{MP} and is quite general for many
types of equations of states of real materials. Even the van der
Waals equation possesses a region around the vapor dome, where
$(\partial^2 V/\partial p^2)_s<0$\cite{vdw}. In 1942 Bethe showed
that {\it all fluids} having sufficiently large specific heats
exhibit smooth loss of convexity of isentropes in the neighborhood
of the saturated vapor line\cite{B}.

The smooth loss of convexity of isentropes creates prerequisites
for a shock wave of rarefaction (SWR) to come into being. Moreover,
other types of anomalous wave structures, such as sonic shocks,
whose Mach number(s) before the shock, after it, or both is (are)
identically equal 1, may be observed in this region and its
vicinity, see Refs.\cite{MP,vdw} for more details. Bearing in mind
all mentioned above, it is highly desirable to have a certain
universal equation, which is not connected with any particular
equation of state but nevertheless can describe hydrodynamics waves
in the discussed region. Despite the apparent importance of the
problem and its long history, much to our surprise, nobody has
payed attention to the following consequence of the fact that
usually the region with the smooth loss of convexity of isentropes
occupies a small domain in $p\,$-$V$ plane. The consequence is that
{\it the corresponding waves are always weak\/} (otherwise the wave
curves\cite{wc} go far away from the domain), and therefore the
problem of the wave description is tractable within the framework
of the macroscopic hydrodynamics in the manner analogous to that
for weak shock waves of compression. To be precise we must
stipulate that we do not consider any singularities related to the
very vicinity to the critical point. In other words, it is implied
the system is far enough from it to avoid any effect of the
singularities.

In the present Letter the mentioned approach is applied to the
problem. As a result the generalized Burgers equation, which
describes hydrodynamic waves in the region, where the derivative
$(\partial^2 V/\partial p^2)_s$ is small and may change sign is
derived. The equation is valid for both signs of the derivative and
therefore describes all the variety of waves, which may exist in
this region. As an example the equation is employed to study steady
shock waves of rarefaction (SWR). The general exact solutions of
the equation describing such shocks is obtained and analysed.

To begin with, let us consider the conventional Burgers equation,
which governs a weak shock wave of compression. In the laboratory
coordinate frame for a wave advancing in the negative direction of
$x$-axis the equation reads\cite{LL}
\begin{equation}\label{CB}
  \frac{\partial p^\prime}{\partial t} -
  c_1\frac{\partial p^\prime}{\partial x} -
  \alpha p^\prime\frac{\partial p^\prime}{\partial x} =
  ac_1^3\frac{\partial^2 p^\prime}{\partial x^2},
\end{equation}
where $p^\prime$ is the pressure variation, so that the pressure in
the wave profile $p$ equals to $p_1 + p^\prime$ (here and in what
follows subscript 1 indicates the value of the corresponding
quantity in the unperturbed medium, i.e., at $x \rightarrow
-\infty$, while subscript 2 will stand for the value of the same
quantity behind a wave, at $x \rightarrow
\infty$), $c(p)$ is the velocity of sound,
%\begin{equation}\label{alpha}
\[  \alpha = \frac{c_1^3}{2V^2}\left(\frac{\partial^2 V}{\partial
p^2}\right)_s,\]
%\end{equation}
$V$ denotes the specific volume ($V \equiv 1/\rho$), $a$ is a
dissipative constant\cite{LL}
%\begin{equation}\label{a}
\[  a \equiv \frac{1}{2\rho_1 c_1^3}\left[\left(\frac{4}{3}\eta
+ \zeta\right) + \kappa\left(\frac{1}{c_v} -
\frac{1}{c_p}\right)\right],\]
%\end{equation}
$\eta$ and $\zeta$ stand for the coefficients of viscosity,
$\kappa$ for the thermal conductivity and $c_{v,p}$ for the
corresponding specific heats.

Eq.~(\ref{CB}) may be regarded as leading approximation to
expansion of a certain more general nonlinear dissipative equation
in powers of both weak nonlinearity and weak dissipation. It is
important that the nonlinear term in this approximation is
considered in non-dissipative limit. Dissipative corrections to it
have additional smallness and may be neglected.

In our case $\left|(\partial^2 V/\partial p^2)_s\right|$ is
anomalously small (it even can vanish). Accordingly, the lowest
nonlinearity $\alpha p^\prime(\partial p^\prime/\partial x)$ may
not correspond to the leading nonlinear term any more. It means
that higher order (in $p^\prime$) terms must be taken into account.
On the other hand, the region, where $(\partial^2 V/\partial
p^2)_s<0$, is narrow, therefore the third derivative $(\partial^3
V/\partial p^3)_s$ also has a certain smallness in this region.
Only the fourth derivative $(\partial^4 V/\partial p^4)_s$ does not
have any smallness there. Thus, at small but finite $p^\prime$
terms $O\left((p^\prime)^2\right), \; O\left((p^\prime)^3\right)$
and $O\left((p^\prime)^4\right)$ may yield contributions of the
same order of magnitude and must be taken into account
simultaneously. It is important that dissipative corrections to the
nonlinear terms remain negligible and all such terms may be
considered in the non-dissipative limit, in the same manner as it
is for the conventional Burgers equation. All the above mentioned
is related exclusively to nonlinear terms. Regarding the
dissipative term on the right hand side of Eq.~(\ref{CB}), there
are no specific reasons for it to be small and no correction to
this term is required.

To derive the desired nonlinear corrections to the left hand side
of Eq.~(\ref{CB}) let us consider an arbitrary adiabatic flow in
the form of a traveling wave. Due to the adiabatic condition the
profile of entropy of such a flow should not change, and since the
state of the medium ahead of the shock front is spatially
homogeneous with $s=const$, we obtain that $s=const$ anywhere.
Then, we may employ the general solution of the continuity and
Euler equations valid for simple waves\cite{LL} and reduce these
equations to the following single equation
\begin{equation}\label{SW}
  \frac{\partial p^\prime}{\partial t} + [v(p^\prime) \pm
  c(p^\prime)]\frac{\partial p^\prime}{\partial x} = 0,
\end{equation}
where the flow velocity $v$ is given by the expression
\begin{equation}\label{v}
  v = \pm \int\limits_{0}^{p^\prime} \frac{dp}{c(p)\rho(p)}.
\end{equation}
Sign plus in Eqs.~(\ref{SW})--(\ref{v}) corresponds to the wave
advancing in the positive $x$ direction, minus to negative.
According to our choice of the direction of the wave propagation in
what follows we take sign minus.

It is important to stress that under the above-specified conditions
the reduction of the set of hydrodynamic equations to
Eqs.~(\ref{SW})--(\ref{v}) is an {\it exact} result valid for {\it
any} nonlinear dependence $c(p)$ and $\rho(p)$, which may be
obtained for the isentropic wave. However, if $p^\prime$ is small
one can expand these functions about the point $p^\prime = 0$. It
is seen straightforwardly that if one truncates the expansion at
zero term, it reduces Eqs.~(\ref{SW})--(\ref{v}) to the linear
acoustic equation. The truncation at the term of order
$O(p^\prime)$ yields the left hand side of Eq.~(\ref{CB}).
Increasing the order of the truncation one can obtain the desired
corrections to the conventional Burgers equation with {\it any}
given accuracy.

Thus, the generalized Burgers equation must have the form
\begin{equation}\label{GB}
  \frac{\partial p^\prime}{\partial t} + u(p^\prime)\frac{\partial
  p^\prime}{\partial x} = c_1^3a\frac{\partial^2 p^\prime}{\partial
  x^2},
\end{equation}
where $u$ stands for the velocity of a point in the wave profile.
For our choice of the propagation direction $u(p^\prime) =
v(p^\prime)-c(p^\prime)$\cite{n1}.

To get the equation in the explicit form one must expand
$u(p^\prime)$ in powers of small $p^\prime$ retaining the terms up
to $O\left((p^\prime)^3\right)$. However, before doing that it is
worth analyzing some general features of Eq.~(\ref{GB}).

First of all let us obtain the general expression for the velocity
$v_s$ of a steady traveling shock, when $p^\prime(x,t) \rightarrow
p^\prime(x+v_st) \equiv p^\prime(\xi)$. In this case Eq.~(\ref{GB})
may be easily integrated. Then, taking into account the boundary
conditions ($p^\prime \rightarrow 0,\; \partial p^\prime/\partial
\xi
\rightarrow 0$ at $\xi\rightarrow -\infty$ and $p^\prime \rightarrow p_2-p_1,\;
\partial p^\prime/\partial \xi \rightarrow 0$ at $\xi\rightarrow
\infty$), we arrive at the following expression for $v_s$
\begin{equation}\label{vs}
  v_s = - \frac{1}{p_2-p_1}\int\limits_{0}^{p_2-p_1}u(p)dp \equiv -\langle
u \rangle,
\end{equation}
where $\langle \ldots \rangle$ denotes average over $p^\prime$ [we
remind that $u = v-c<0$ see also Eq.~(\ref{v})].

Let us show now that $v_s$ given by Eq.~(\ref{vs}) for SWR
satisfies the evolutionary conditions. The conditions say that in
the co-moving coordinate frame the velocity of the flow before the
shock should be bigger than $c_1$, while the velocity behind the
shock should be smaller than $c_2$, i.e., $M_1>1>M_2>0$, where
$M_{1,2}$ stand for the corresponding Mach numbers. In the
laboratory coordinate frame the medium before the shock is in the
rest state with zero velocity. Accordingly, in the co-moving
coordinate frame the flow velocity here is $v_s$. Thus, the first
evolutionary condition says $v_s > c_1$. Behind the shock the flow
velocity in the laboratory frame is $v_2$. Consequently, the second
condition yields $v_s+v_2<c_2$. Bearing in mind Eq.~(\ref{vs}) both
the conditions may be written as follows
\begin{equation}\label{evol}
  c_1 \equiv - u_1< - \langle u \rangle < c_2-v_2 \equiv -u_2
\end{equation}
Finally, taking into account that for a simple wave and the chosen
direction of propagation the sign of derivative $du/dp^\prime$ is
opposite to that for $(\partial^2 V/\partial p^2)_s$\cite{LL},
i.e., for negative $(\partial^2 V/\partial p^2)_s$ function
$-u(p^\prime)$ monotonically increases with decrease of $p^\prime$,
and that for SWR $p_2$ is smaller than $p_1$, we reduce inequality
(\ref{evol}) to evident.

Let us proceed with the derivation of the generalized Burgers
equation. From all the above-mentioned it is clear that in the
discussed region the derivative $(\partial^2 V/\partial p^2)_s$ may
be approximated as follows
\begin{equation}\label{der}
(\partial^2 V/\partial p^2)_s \approx
\frac{1}{\rho_1}\frac{1}{(\rho_1c_1^2)^2}\left[- \epsilon^2 +
\frac{\mu^6}{4}\left(\frac{p-p_m}{\rho_1c_1^2}\right)^2\right].
\end{equation}
Here $p_m$ is the value of $p$, corresponding to the local minimum
of the derivative, $\epsilon$ is a small dimensionless quantity
$\mu = O(1)$ and power $\mu^6$ as well as the numerical coefficient
are introduced for convenience of further notations.

According to Eq~(\ref{der}), $(\partial^2 V/\partial p^2)_s<0$ at
$p_m-\Delta < p < p_m +\Delta$, where $\Delta =
2\rho_1c_1^2\epsilon/\mu^3 = O(\epsilon)$. Then, expanding
$u(p^\prime)$ in powers of $p^\prime$, taking into account that for
the problem in question $p^\prime$ is of order $\epsilon$, or
smaller and dropping terms higher than $O(\epsilon^3)$, after some
calculations we reduce the general equation (\ref{GB}) to the form
of Eq.~(\ref{CB}), where term $\alpha p^\prime$ should be replaced
by the following expression
%\begin{equation}\label{alpha1}
\begin{eqnarray*}%\label{alpha1}
  \alpha p^\prime & \rightarrow &
  \frac{\rho_1^2c_1^3}{2}\left[\left(\frac{\partial^2V}{\partial
  p^2}\right)_sp^\prime + \frac{1}{2}\left(\frac{\partial^3V}{\partial
  p^3}\right)_sp^{\prime 2}\right.\\
  & & \left.+ \frac{1}{6}\left(\frac{\partial^4V}{\partial
  p^4}\right)_sp^{\prime 3}\right],
\end{eqnarray*}
and all derivatives $(\partial^n V/\partial p^n)_s$ are taken at
$p=p_1.$

It is convenient to rewrite the equation in more universal
dimensionless form. Let us introduce new variables

\[ y \equiv \frac{p^\prime}{\Delta};\;\; \tau \equiv
\frac{\epsilon^6}{\mu^6c_1a}t;\;\; \zeta \equiv
\frac{\epsilon^3}{\mu^3c_1^2a}( x +
c_1t). \]

In these variables involving expression (\ref{der}) one can reduce
the generalized Burgers equation to the following final form
\begin{equation}\label{final}
  \frac{\partial y}{\partial \tau} - f(y)\frac{\partial
  y}{\partial \zeta} = \frac{\partial^2 y}{\partial \zeta^2},
\end{equation}
where
\begin{equation}\label{f}
  f(y) \equiv (z^2 -1)y + zy^2 + \frac{y^3}{3};\;\; z \equiv
  \frac{p_1 - p_m}{\Delta}.
\end{equation}

In what follows we examine solutions of this equation, which
correspond to SWR. In this case we should supplement
Eq.~(\ref{final}) with the boundary conditions $\partial y/\partial
\zeta \rightarrow 0$ at $\zeta \rightarrow \pm \infty$; $y
\rightarrow y_1 = 0$ at $\zeta
\rightarrow -\infty$; $y
\rightarrow y_2 = const<0$ at $\zeta \rightarrow \infty$ and bear
in mind that $y\leq 0$; $-1\leq z \leq 1.$ To study steady SWR it
is convenient to introduce a traveling coordinate $\eta \equiv \
\zeta + \nu
\tau$, where in accordance with Eq~(\ref{vs})
the dimensionless velocity $\nu$ equals the following expression
%\begin{equation}\label{nu}
\[  \nu = \frac{z^2-1}{2}y_2 + \frac{zy_2^2}{3} +
\frac{y_2^3}{12}\]
%\end{equation}
Then, integration of Eq.~(\ref{final}) yields
\begin{eqnarray} \label{int}
\lefteqn{
  \eta = -12\left[ - \frac{\ln|y|}{y_2y_3y_4} +
  \frac{\ln|y-y_2|}{y_2(y_3-y_2)(y_4-y_2)} \right.}
  \nonumber\\
  & & + \left.\frac{\ln|y-y_3|}{y_3(y_3-y_4)(y_3-y_2)} +
  \frac{\ln|y-y_4|}{y_4(y_4-y_3)(y_4-y_2)} \right],
\end{eqnarray}
where $y_{3,4}$ are the roots of the equation
\begin{equation}\label{pol}
  y^2 + (4z + y_2)y + y_2^2 + 4zy_2 + 6(z^2-1) = 0
\end{equation}

Let us show that $y_{3,4}$ are always real and one of these root is
smaller than $y_2$, while the other is bigger than $y_1$ (we remind
that by definition $y_1=0$). In other words the left hand side of
Eq.~(\ref{pol}) is negative at $y_2<y<0$ and any possible values of
$z$ and $y_2$. To prove it note that for SWR pressure $p_2$ must
satisfy obvious conditions $p_m - \Delta
\leq p_2 \leq p_1$, which in the dimensionless variables are
transformed into the following inequalities
%\begin{equation}\label{y2}
\[  -(z+1) \leq y_2 \leq 0.\]
%\end{equation}

Due to the fact that $y^2$ enters into Eq.~(\ref{pol}) with a
positive coefficient to prove negativeness of this polynomial at
$y_2<y<0$ it suffices to examine its values at the marginal points
$y=y_{1,2}$. At $y=y_1=0$ we obtain
\begin{equation}\label{pol1}
y_2^2 + 4zy_2 + 6(z^2-1).
\end{equation}
For the same reason it is sufficient to inspect the values of
polynomial (\ref{pol1}) at the marginal values of $y_2$, namely at
$y_2 = -(z+1),\; y_2 = 0$. It is straightforward to see that the
marginal values of Eq.~(\ref{pol1}) are negative at $|z|<1$.
Negativeness of polynomial (\ref{pol}) at $y=y_2$ is proved in the
same manner. The proved relative position of points $y_{1,2,3,4}$
guarantees that the derivative $dy/d\eta$ is negative at
$y_2<y<y_1$ for any possible values of $z$ and $y_2$, i.e., the
profile of the steady SWR is a monotonically decreasing function of
$\eta$.

To end this Letter we present several particular versions of
general solution (\ref{int}), when the dependence $y(\eta)$ may be
obtained explicitly.

(i) $z=1 \; y_2 =-2$, which corresponds the maximal possible
amplitude of SWR ($p_1 = p_m + \Delta, \; p_2 = p_m - \Delta)$.

\[ \eta = -\frac{3}{2}\ln \left|\frac{2+y}{y}\right| -
\frac{3}{2\sqrt{5}}\ln \frac{y+1+\sqrt{5}}{\sqrt{5}-1-y}. \]

(ii) $z=1\; \left|y_2\right| \ll 1\; p_1 = p_m + \Delta, \; p_1-p_2
\ll \Delta.$ In this case the leading approximation in small
$\left|y_2\right|$ yields

\[ y = y_2\sqrt{\frac{1 + \tanh (y_2^2\eta/3)}{2}} \]

(iii) $z=-(1+y_2),\; \left|y_2\right| \ll 1\; [p_1=p_m-\Delta -
(p_1-p_2),\; p_1-p_2 \ll \Delta$]. The leading approximation in
$\left|y_2\right|$ in this case results in the following profile

\[ y = y_2\left[1-\sqrt{\frac{1-\tanh{(y_2^2 \eta /3)}}{2}}\;\right] \]

Note, that while in the present Letter only steady solutions of
Eq.~(\ref{final})--(\ref{f}) corresponding to SWR are discussed,
the equation itself describes much more broad spectrum of problems
of steady and non-steady flows, including such very interesting
issues, as sonic waves, evolution of arbitrary initial profiles,
collision of shocks, etc. The corresponding study is in progress
and results will be reported elsewhere.

One of the authors (M.I.T.) is grateful to Peter Fulde for
invitation to MPI-PKS, which provides the opportunity to complete
the present study, and to the entire staff of the Institute for
kind hospitality. This work was supported by the Grant-in-Aid for
Scientific Research (No. 11837006) from the Ministry of Education,
Culture, Sports, Science and Technology (Japan).

% now the references. delete or change fake bibitem. delete next three
%   lines and directly read in your .bbl file if you use bibtex.

% \begin{figure}
% \caption{}
% \label{}
% \end{figure}

% \begin{table}
% \caption{}
% \label{}
% \begin{tabular}{}
% \end{tabular}
% \end{table}

\end{twocolumns}

\begin{references}
\bibitem{wc}A curve, which consists of states connecting
initial and final states by a given wave solution.
\bibitem{n1} For a wave traveling in the opposite direction we
would have $u(p^\prime)=v(p^\prime)+c(p^\prime).$
\end{references}
\end{document}